\documentclass[]{aa}  

\usepackage{graphicx}
\usepackage{txfonts}
\usepackage{xcolor,colortbl}
\usepackage{hyperref}
\usepackage{multirow}
\usepackage[normalem]{ulem}
\usepackage{txfonts}
\usepackage{upgreek}
\makeatletter
\renewcommand*\aa@pageof{, page \thepage{} of \pageref*{LastPage}}
\makeatother
\newcommand{\Msun}{{\rm\, M_\odot}}

\newcommand{\pc}{{\rm\, pc}}

\begin{document}

   \title{Chemical analysis of the Milky Way's Nuclear Star Cluster}
\subtitle{Evidence for a metallicity gradient}
   \author{M. Schultheis
            \inst{1}
            \and
            L. Serrano
            \inst{1,2}
            \and
            B. Thorsbro
            \inst{1}
            \and
            F. Nogueras-Lara
            \inst{3}
            \and
            A. Feldmeier-Krause
            \inst{4}
            \and
            G. Nandakumar
            \inst{5}
            \and
            K. Fiteni
            \inst{6,7}
            \and
            M.C. Sormani
            \inst{6}
    }
   \institute{Université Côte d’Azur, Observatoire de la Côte d’Azur, Laboratoire Lagrange, CNRS, Blvd de l’Observatoire, 06304 Nice, France\\
    \email{mathias.schultheis@oca.eu}
    \and
    MAUCA – Master track in Astrophysics, Université Côte d’Azur, Observatoire de la Côte d’Azur, Parc Valrose, 06100 Nice, France
    \and
    Instituto de Astrof\'isica de Andaluc\'ia (CSIC), Glorieta de la Astronom\'ia s/n, E-18008 Granada, Spain
    \and
    Department of Astrophysics, University of Vienna, T\"urkenschanzstrasse 17, 1180 Wien, Austria
    \and
    Aryabhatta Research Institute of Observational Sciences, Manora Peak, Nainital 263002, India
    \and
    Como Lake centre for AstroPhysics (CLAP), DiSAT, Università dell’Insubria, Via Valleggio 11, 22100 Como, Italy
    \and
    Institute of Space Sciences \& Astronomy, University of Malta, Msida MSD 2080, Malta
    }

   \date{Received Month Day, Year; accepted Month Day, Year}

\abstract{The Milky Way nuclear star cluster (MWNSC) is located together with its surrounding nuclear stellar disc (MWNSD) in the Galactic centre and they dominate the gravitational potential within the inner 300\,pc. However, the formation and evolution of both systems and their possible connections are still under debate.}{We reanalyse the low-resolution KMOS spectra in the MWNSC  with the aim to improve the stellar parameters ($\rm T_{eff}$, $\rm \log\,g$, and $\rm [M/H])$ for the MWNSC.}{ We use an improved line-list, especially dedicated for cool M giants allowing to improve the stellar parameters and to obtain in addition global $\rm \alpha$-elements. A comparison with high-resolution IR spectra (IGRINS) gives very satisfactory results pinning down the uncertainties to $\rm T_{eff} \simeq 150\,K$, $\rm log\,g \simeq  0.4\,dex$, and $\rm [M/H] \simeq 0.2\,dex$. Our $\rm \alpha$-elements agree within 0.1\,dex compared to the IGRINS spectra.}{We obtain a high-quality sample of 1140 M giant stars where we see an important contribution of a metal-poor population ($\rm \sim 20\,\%$) centered at $\rm [M/H] \simeq -0.7\,dex$ while the most dominant part comes from the metal-rich population with $\rm [M/H] \simeq 0.26\,dex$. We construct a metallicity map and find a metallicity gradient  of $\rm \sim -0.1 \pm 0.02 \,dex/pc$ favouring the inside-out formation scenario for the MWNSC.} {} 


   \maketitle

\section{Introduction}

Nuclear star clusters (NSCs) are compact and massive clusters located at the dynamical centres of their host galaxies. Their effective radii, within which half of the cluster light is contained, are found to span a wide range from $0.4$ to $44$ pc \citep{Neumayer2020}. The two proposed formation mechanisms are first the inspiral and mergers of massive star clusters (e.g. \citealt{Antonini2012}, \citealt{Hartmann2011}), and second in-situ formation (e.g. \citealt{Aharon2015},\citealt{Brown:2018}). However, both scenarios could possibly contribute to the growth of NSCs \citep{Boker:2010}. Moreover, the in situ formation can be caused by several secular mechanisms such as bar-driven gas infall \citep{Shlosman:1990}, dissipative nucleation \citep{Bekki:2006,Bekki:2007}, tidal compression \citep{EmsellemVandeVen:2008} or magneto-rotational instability \citep{Milosavljevic:2004}. \citet{Fahrion2021} argued that there appears to be a distinction of the formation mechanism  with galaxy mass: Less massive NSCs are produced mainly by cluster infall, while  more massive NSCs are formed in-situ with a transition galaxy mass of about  $\rm \sim \, 10^{9} M_{\odot}$. \\

The Milky Way also features a NSC (hereafter called MWNSC) in its centre with an effective radius between $\rm 4.2\pm0.4\ pc$ and $\rm 7.2\pm2.0\ pc$ (\citealt{Schoedel:2014}, \citealt{Fritz:2016}, \citealt{Gallego-Cano2020}). The mass of the MWNSC lies between $2.1\pm0.7\times10^{7}$ and $4.2\pm1.1\times10^{7}\ M_{\odot}$ (\citealt{Schoedel:2014,Schoedel:2020}, \citealt{Chatzopoulos:2015}, \citealt{Feldmeier-Krause2014}, \citealt{Feldmeier:2017b}).\\

The MWNSC is embedded inside the Milky Way nuclear stellar disc (MWNSD), which is   a central, kinematically cold, distinct, rotating structure ($\rm v/ \sigma \simeq 1$, \citealt{review_paper}). The MWNSD can be described by an exponential intensity  radial profile and is thought to be formed from gas brought to the centre by the bar via the Central Molecular Zone  (see e.g. \citealt{Sormani2024}).  The  radial and vertical scale-lengths  for the MWNSD are  $R = 88.6^{+9.2}_{-6.9} \pc$ and $H=28.4^{+5.5}_{-5.5} \pc$ respectively (\citealt{Gallego-Cano2020}, \citealt{Sormani2022}) , and a total mass $M_{\rm NSD} = 10.5^{+1.1}_{-1.0} \times10^8 \,\Msun$ (\citealt{Sormani2022}).

\begin{figure*}[!htbp]
\centering
\includegraphics[width=0.9\textwidth]{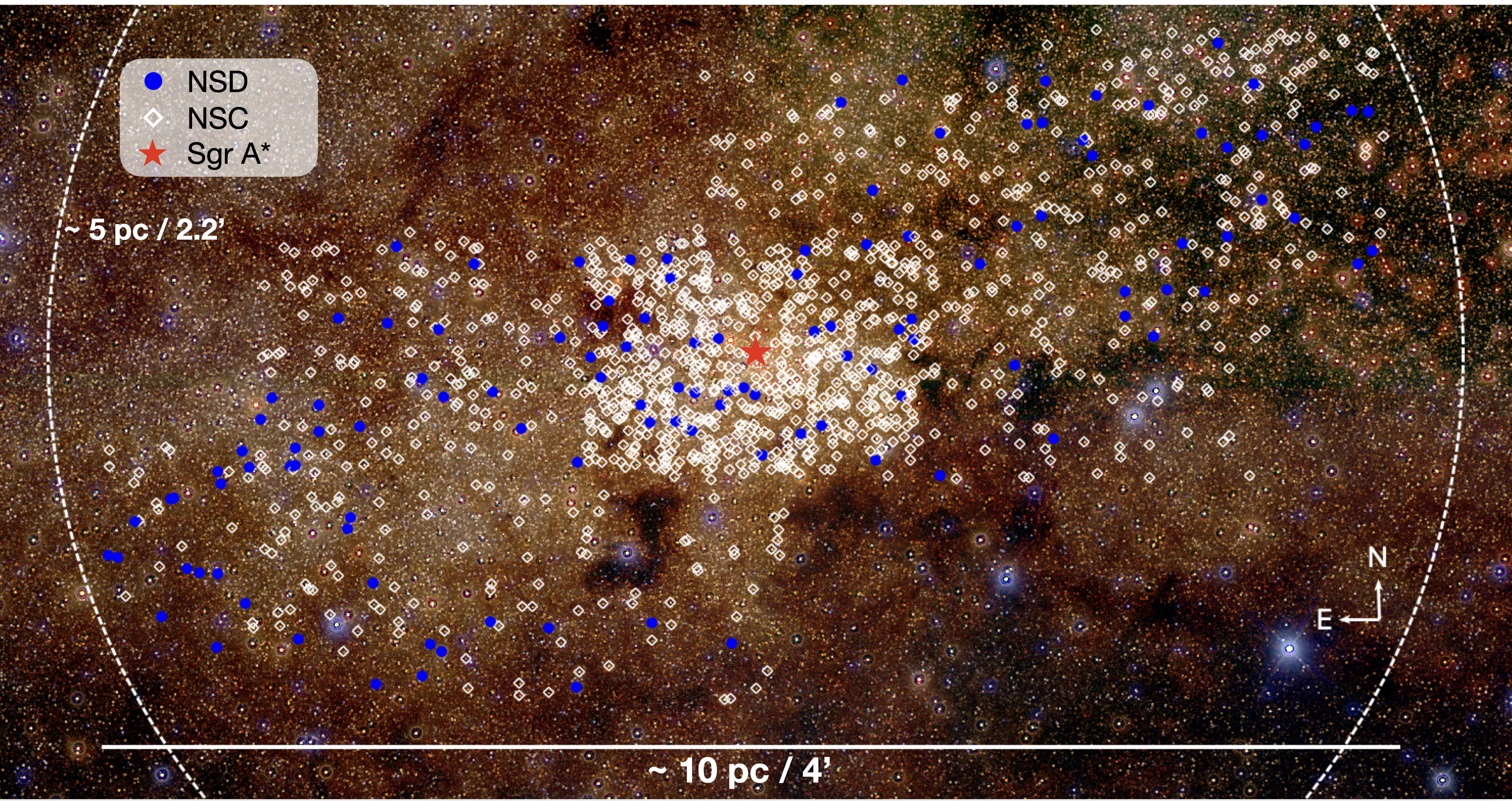}
\caption{GALACTICNUCLEUS $JHK_s$ false-colour image of the region covered by FK1720. Blue circles and white diamonds mark stars classified as belonging to the MWNSD and MWNSC, respectively, according to the criterion described in Sect.\,\ref{membership}. The white dashed line indicates the effective radius of the MWNSC, and the compass shows Galactic coordinates.}
\label{scheme}
\end{figure*}

While several photometric studies suggest that the MWNSD and MWNSC may have undergone distinct formation histories \citep{Nogueras-Lara:2020a,Nogueras-Lara:2021apj}, \cite{Nogueras-Lara2023} found kinematic and metallicity gradients that would suggest a smooth transition between both components. They propose that the MWNSD and MWNSC might be part of the same structure. Also, \cite{Seth:2006} found discs or rings superimposed onto the observed NSCs in the nuclear regions of 14 nearby galaxies. They suggest that their observations may support an in situ formation where NSCs are formed from stars losing angular momentum from the NSDs in which NSCs are embedded. 
The links between NSDs and NSCs are however still under debate (see also \citealt{review_paper}).   \\

Due to the extreme extinction towards the MWNSC \mbox{(${\rm A_{V} \sim 30}$)}, only observations in the infrared are possible (\citealt{Schoedel2010}, \citealt{Nogueras-Lara2018}). Even spectroscopic surveys such as APOGEE could only observe a few very  luminous stars in the NSC due to its limiting sensitivity (\citealt{Schultheis2020}). 
\citet{Feldmeier-Krause2017} and \citet{Feldmeier-Krause2020} used an extensive dataset 
of KMOS data (ESO/VLT) to perform a chemical study (i.e. metallicities) of stars located in the MWNSC. They find a predominantly metal-rich population with a mean metallicity of 0.34\,dex with  some indications of an anisotropic metallicity distribution function, i.e. a higher fraction of sub-solar metallicity stars in the Galactic North. They argue that  this anisotropy could be due to star cluster infall events (see e.g. \citealt{Antonini2012}, \citealt{Perets2014}, \citealt{ArcaSedda2020}, \citealt{Do2020}). \citet{Feldmeier-Krause2025} conducted a first spectroscopic survey from the MWNSC to the inner MWNSD, out to $\rm \pm 32\,pc$ from 
Sgr~A$^*$. They provided a first global map of the mean metallicity and found a decrease of $\rm [M/H]$ towards the centre of the MWNSC. However, as discussed by the authors, this decrease in metallicity could be  due to a projection effect.

One difficulty in all the above mentioned studies is the fact that these spectra have low spectral resolution ($\rm R \sim 3000-4000$). The vast majority of the studied stellar populations are cool M giant stars  with temperatures below 4000\,K. The spectral analysis  for these stars is extremely challenging, even for high-resolution spectra (see e.g. \citealt{Thorsbro2023}, \citealt{Ryde2025}, \citealt{Nandakumar2025}) where line blends with molecules is a big issue. 

In this paper we re-analyse the KMOS dataset of \citet{Feldmeier-Krause2020} taking advantage of a recent improved line list, specifically adapted for cool M giants and which has been also used for high-resolution spectroscopic studies (\citealt{Ryde2025}). We use available high-resolution spectra of cool M giants covering a similar stellar parameter space to validate our method.

\section{Data} \label{Data}

We use the dataset of \citet{Feldmeier-Krause2017} and \citet{Feldmeier-Krause2020} (hereafter referred to as FK1720) observed by KMOS (\citealt{KMOS}) at the VLT which we have re-analysed. KMOS consists of 24 IFUs with a field of view of $ \rm 2.8\arcsec \times 2.8\arcsec$ each. 
 The spatial scale is $\rm 0.2\arcsec/pix \times 0.2\arcsec/pix$. The central field was observed twice by \citet{Feldmeier-Krause2017} and centered on the MWNSC. We refer here to \citet{Feldmeier-Krause2017} for more details of the dataset. \citet{Feldmeier-Krause2020} extended the central field by observing six mosaic fields within the half-light radius of the MWNSC ($\sim$ 4-5\,pc; \citealt{Schoedel2014, Gallego-Cano2020}).  Figure\,\ref{scheme} shows the target field and the observed stars.

 The wavelength coverage of these spectra is between 19340\,\AA -- 24600\,\AA\, with a spectral resolution of $\rm\sim~2.8\,\AA\, pixel^{-1}$. As the spectral resolution of KMOS varies spatially, we used the resolution maps  as well as the radial velocities from FK1720 as an input for the Bayesian STARKIT code (\citealt{starkit}). We refer for a more detailed description of the dataset to FK1720. Most of the stars have repeated observations which we treat separately.

\subsection{MWNSC membership} \label{membership}

To identify MWNSC stars and obtain a photometric counterpart for each target star, we cross-matched the FK1720 KMOS stellar sample with the $\rm HK_s$ photometry from the GALACTICNUCLEUS survey \citep{Nogueras-Lara2018,GALACTICNUCLEUS}. This catalogue provides state-of-the-art, high-angular-resolution ($\sim0.2''$) near-infrared photometry of the Galactic centre. To avoid saturation for GALACTICNUCLEUS sources with $K_s\lesssim11.5$, we complemented the data with the SIRIUS/IRSF catalogue \citep[e.g.][]{Nagayama2003,Nishiyama2006}, replacing saturated stars with undetected ones as described in \citet{Nogueras-Lara:2022a}.

We then used common stars to align both catalogues and adopted a maximum search radius of $\sim 0.3''$, corresponding to half the typical angular resolution of the KMOS data \citep[e.g.][]{Feldmeier-Krause2017}. This yielded $\sim1300$ stars with $\rm HK_s$ photometry.

\begin{figure}[!htbp]
\centering
\includegraphics[width=0.49\textwidth]{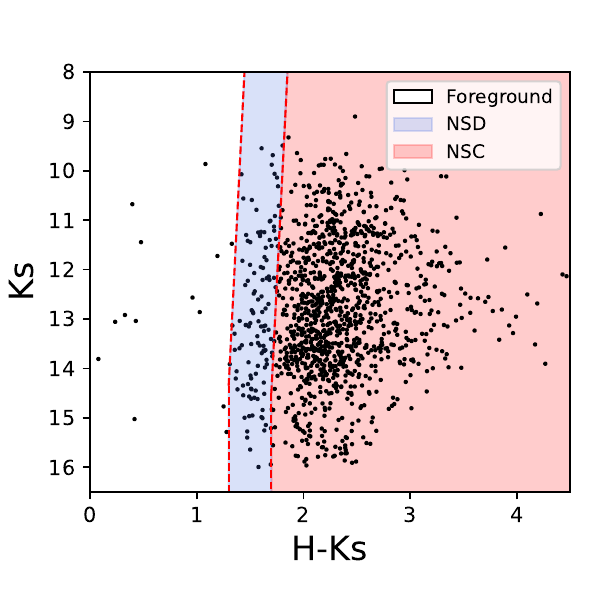}
\caption{H--K vs. K colour magnitude diagram of our data sample. Stars member of the MWNSC are indicated in red, in blue stars of being member of the MWNSD are indicated and in white we show the foreground objects. For our work here we only use stars in the MWNSC.}
\label{CMD}
\end{figure}

To separate Galactic components, we applied a photometric criterion to distinguish between foreground stars (likely belonging to spiral arms along the line of sight and to the Galactic bulge; e.g. \citealt{Nogueras-Lara:2021b}), MWNSD stars, and MWNSC stars, following the method described in \citet{Nogueras-Lara2023}. This approach assumes a correlation between distance and extinction along the line of sight, which allows us to statistically differentiate these components by applying a colour cut, as shown in Fig.\,\ref{CMD}.

\subsection{Data completeness}

\begin{figure}[!htbp]
\centering
\includegraphics[width=0.49\textwidth]{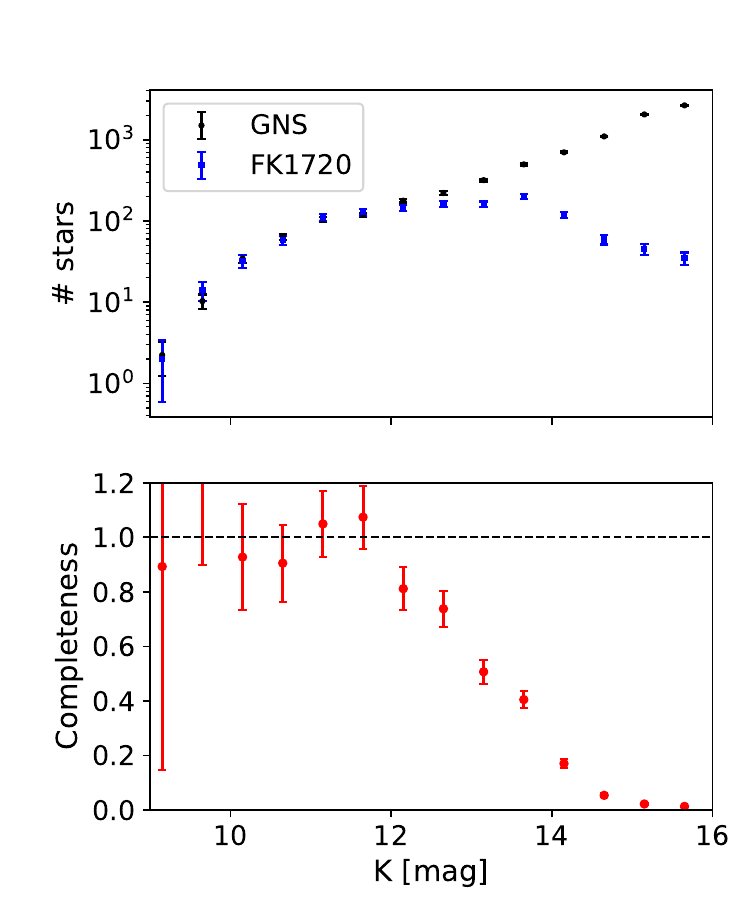}
\caption{ Completeness of the KMOS FK1720 sample. Upper panel: $\rm K_s$ luminosity functions from the KMOS sample and the reference sample of GALACTICNUCLEUS (GNS) stars in the region. The associated uncertainty was estimated as the square root of the number of stars per magnitude bin. Lower panel: Completeness function obtained by comparing the two $\rm K_s$ luminosity functions.}
\label{completeness}
\end{figure}

To estimate the completeness of the KMOS sample, we constructed a $\rm K_s$ luminosity function and compared it with a scaled reference function based on all GALACTICNUCLEUS stars detected in $\rm K_s$ within the target region. We then computed the completeness by assuming that the reference catalogue is approximately fully complete in the relevant region and magnitude range \citep[e.g.][]{Nogueras-Lara:2020a}. Figure\,\ref{completeness} shows the resulting completeness, indicating that the spectroscopic sample is $\gtrsim60\%$ complete up to $\rm K_s=13$.

\subsection{Analysis} \label{Analysis}
As in FK1720,  we  have used the full spectral fitting code STARKIT  (\citealt{starkit}),   but added some specific improvements for the analysis of cool stars. STARKIT interpolates the template spectra and applies a Bayesian sampling (Multinest, \citealt{multinest}) to obtain the best spectral fit using a grid of synthetic spectra. 
A key improvement in this work is the use of a dedicated synthetic model grid optimised for cool giants (see Sect.~\ref{SMEmodel}). By contrast, FK1720 based their analysis on the PHOENIX stellar library (\citealt{Phoenix2013}), which neither incorporates the updated line lists adopted here nor includes support for NLTE effects.

In addition, we have performed for each of the spectra  a continuum normalization using a sixth order polynomial function  and applying  a two-sigma clipping. As we assume that all stars in our sample are M giants, we set a uniform prior in temperature ($\rm 2800 < T_{eff} < 4600\,K$) and in the surface gravity ($\rm -0.5 < log\,g <3$), typical values for M giants. Contrary to FK1720 we do not use any photometric criteria to constrain log\,g due to the large uncertainties in the photometric surface gravities.
Our specific grid also allows to compute $\rm \alpha$-elements (see Sect.~\ref{alpha}). 
Our dedicated synthetic model grid covers the following stellar parameter ranges: $\rm 2800 < T_{eff} < 4600\,K$, $\rm -0.5 < log\,g <3$, $\rm -1.2 < \rm [M/H] < 0.6$, and $\rm -0.4 < \rm [\alpha/Fe] < 0.6$.

Each observed spectrum was modeled using synthetic spectra {from a  Spectroscopy Made Easy \citep[SME,][]{sme_code,sme_code_new}-based grid convolved to match the instrument resolution. The spectral grid spans stellar atmospheric parameters including effective temperature ($\rm T_{eff}$), surface gravity (log\,g), metallicity ([M/H]), and $\rm alpha$-enhancement ([$\alpha$/Fe]). The observational data were preprocessed to exclude regions contaminated by $\rm Br_{\gamma}$, $\rm NaI$, $\rm CaI$, and the CO molecular features, in a similar way as in FK1720 (see Fig.~\ref{Example_Starkit}).
In this work, we  also derive the $\rm \alpha$-abundances of our stars (see Sect.~\ref{SMEmodel}).

A Bayesian inference framework was applied using the MultiNest nested sampling algorithm to explore the posterior distribution of stellar parameters. Likelihood evaluation was based on a $\chi^{2} $ comparison between the observed and model spectra.

\begin{figure}[!htbp]
\centering
\includegraphics[width=0.49\textwidth]{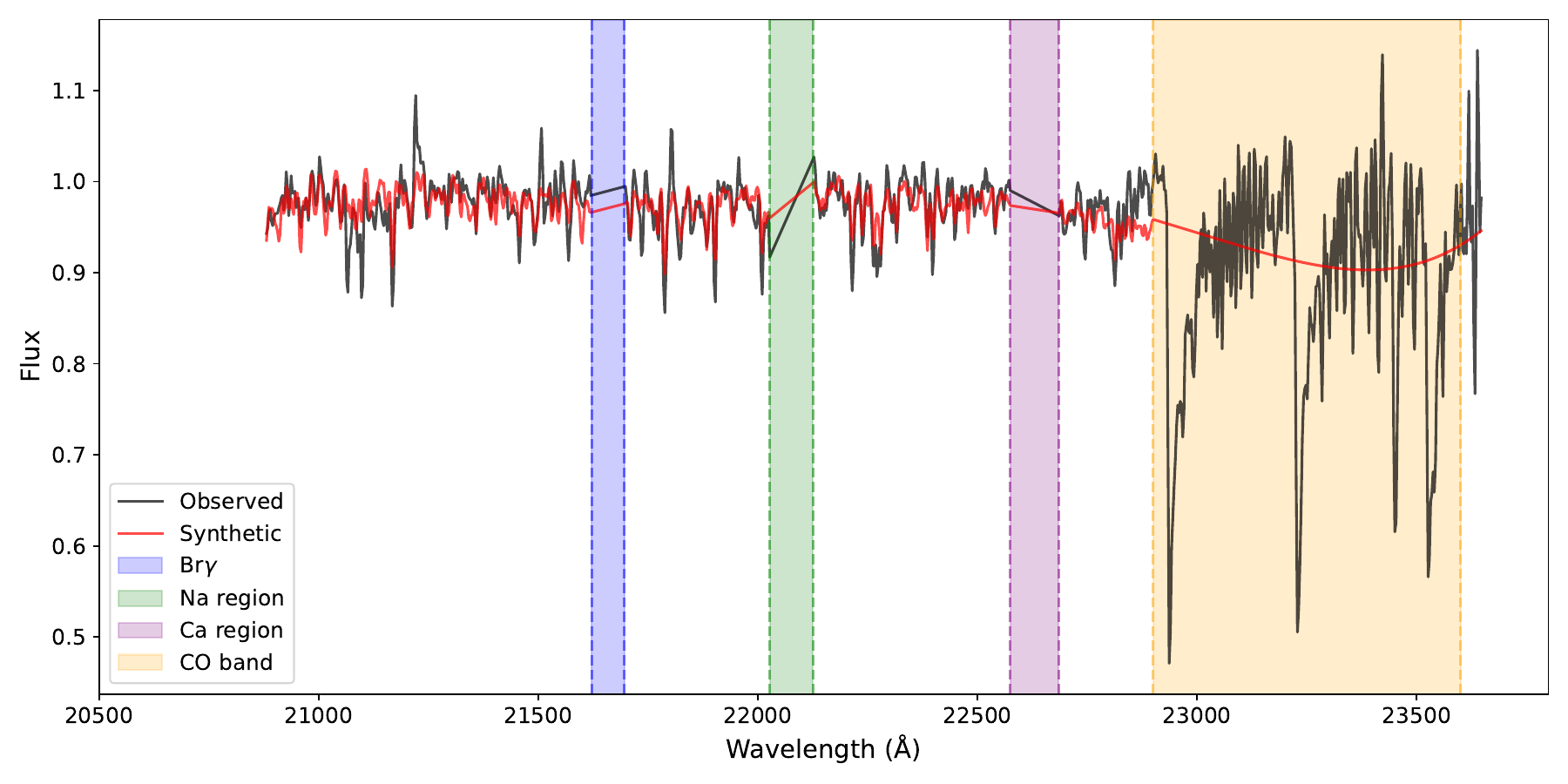}
\caption{Example of a spectral fit using STARKIT for starId 30001001. The normalized, observed spectrum is in black, the best synthetic fit is indicated in red. The vertical bands show the masks applied: Blue is the $\rm Br_{\gamma}$ region, green the region around the NaI line, magenta the region around the CaI line and orange the CO band heads}
\label{Example_Starkit}
\end{figure}

The posterior probability distributions for each stellar parameter were obtained using the MultiNest algorithm, which provides a statistically rigorous sampling of the multidimensional parameter space. The sampling yields posterior samples that encode both the likelihood $\rm \chi^{2}$-based spectral fit and the imposed priors. For each parameter, we derive the 50th percentile (median) of the posterior distribution together with its standard deviation (the square root of the variance in the posterior). To complement the posterior statistics, we compute the reduced $\rm \chi^{2}$ of the best-fit model, defined as:
\begin{equation*}
\chi^2_\nu = \frac{1}{N - p} \sum_{i=1}^{N} 
\left( \frac{f_{\mathrm{obs},i} - f_{\mathrm{mod},i}}{\sigma_i} \right)^{2},
\end{equation*}

\noindent
where N is the number of spectral data points, p is the number of free parameters, $\rm f_{\rm{obs},i}$ and $\rm f_{\rm{mod},i}$ are the observed and model fluxes, and $\rm \sigma_{i}$  is the per-pixel uncertainty. 

Special attention is given to cases where posterior distributions are truncated near the
grid edges, particularly for $\rm T_{eff}$, $\rm log\,g$, and $\rm [M/H]$, where synthetic spectra are not defined outside the grid limits. In such cases, we flag these spectra and do not consider them for further analysis.

The $\alpha$-elements have been obtained by running STARKIT in a second time fixing there the stellar parameters $\rm T_{eff}$ and $\rm log\,g$ but using [M/H] and $\rm [\alpha/Fe]$ as a free parameter. As in the first step we use all the posterior parameters to trace the uncertainties as well as the border flags. For our prior we assume that the $\alpha$- elements follow the Galactic prior, meaning that for  sub-solar metallicities, the $\rm \alpha-elements$ are enhanced while for above-solar metallicities, $\rm \alpha$-elements are solar or sub-solar.

Figure~\ref{Example_Starkit} shows a typical spectrum of an M giant, in this case it is the starId 30001001 with $\rm T_{eff}=3550 \pm 90\,K$, $\rm log\,g = 1.1 \pm 0.18\,dex$, $\rm [M/H] = 0.01 \pm 0.15\,dex$, and        $\rm [\alpha/Fe] =-0.02 \pm 0.11\,dex$.

\subsection{New Model grid} \label{SMEmodel}

To determine the stellar parameters --- effective temperature ($T_{\mathrm{eff}}$), surface gravity ($\log g$), and metallicity ([M/H]) --- together with the $\upalpha$-element abundances of our targets, we generate a grid of synthetic spectra using the spectral synthesis code SME \citep[][]{sme_code,sme_code_new}. SME interpolates on a grid of one-dimensional (1D) MARCS atmosphere models \citep{marcs:08}, which are hydrostatic, spherically symmetric models computed under the assumptions of chemical equilibrium, homogeneity, and total flux conservation (radiative plus convective, with convective flux calculated via a mixing-length prescription).

In order to synthesize the spectra, an accurate list of atomic and molecular energy level transitions is required. In the list of atomic energy level transitions, we used the solar centre intensity atlas \citep{solar_IR_atlas} to update wavelengths and line strengths (astrophysical $\log gf$-values) \citep{thorsbro:17,Nandakumar:2024}. Since molecular lines are strong features in our spectra, the adopted line list includes relevant molecular transitions. For CN—the most dominant molecule apart from CO, whose lines dominate in the 2.3\,$\upmu$m bandhead region—we use the list of \citet{sneden:14}. The CO line data are from \citet{Li:2015}. At shorter wavelengths of our spectral region, SiO, H$_2$O, and OH are important; their line lists are taken from \citet{langhoff:07}, \citet{Polyansky:2018}, and \citet{brooke:16}, respectively.

The spectrum synthesis was carried out without assuming local thermodynamic equilibrium (LTE). Instead, we used pre-computed grids of departure coefficients, $b_{i} = n_{i}/n^{*}_{i}$, where $i$ denotes the level index for NLTE and LTE populations $n$ and $n^{*}$, respectively. These coefficients were used to correct the LTE line opacities following the method described in Section~3 of \citet{sme_code_new}. For magnesium, silicon, and calcium, the grids of departure coefficients are those described in \citet{amarsi:grids}, with model atoms from \citet{osorio:mg} and \citet{osorio:ca} for magnesium and calcium (fine structure collapsed) and from \citet{amarsi:si} for silicon. For iron, we use the grid of departure coefficients and model atom presented in \citet{amarsi:fe_new}.

The resulting SME model grid, spanning a range of $T_{\mathrm{eff}}$, $\log g$, [M/H], and [$\upalpha$/Fe] values, is then used in the STARKIT framework \citep{starkit,starkit_2}, which employs the MultiNest algorithm \citep{multinest} to perform a probabilistic grid search. STARKIT interpolates within the synthetic grid to find the model spectra that best match the observations, yielding posterior probability distributions for the stellar parameters and $\upalpha$-element abundances. This combined approach allows us to systematically explore the parameter space, account for molecular and NLTE effects in our modeling, and derive robust values for $T_{\mathrm{eff}}$, $\log g$, [Fe/H], and [$\upalpha$/Fe].

\subsection{Quality control}
 Most of our sample has multiple exposures and we decided to keep only stars with $\rm\Delta T_{eff} < 500 \,K$, $\rm \Delta log\,g < 0.5\,dex$, and $\rm \Delta [M/H] < 0.3\,dex$. We only kept sources with a $\rm SNR > 10$ and rejected sources which are within $\rm 5\%$ to the border grid of our grid of synthetic spectra.
 As pointed out by \citet{Feldmeier-Krause2017}, fringes can occur in the stellar spectra if the star is located at the edge of the IFU. As most of the stars had two or more exposures, we kept only those stars which fulfill the criteria mentioned above and  took the mean of the stellar parameters. In total, our sample consists of 1140 stars. As mentioned in Sect.~\ref{membership}, we exclude foreground stars and stars located in the NSD.

\section{Validation with high-resolution spectra} \label{Validation}

We take advantage of the high-resolution, high signal-to-noise sample of cool M giants in the solar neighbourhood from \citet{Nandakumar2024}, hereafter referred to as the SN comparison sample. These stars have been observed with the IGRINS spectrograph on the Gemini South telescope covering the full H- and K- bands with a spectral resolution of $\rm R = 45.000$. Their sample consists of 30 cool M giants with $\rm T_{eff} \leq 4000\,K$ and a high signal-to-noise ratio ($\rm SNR > 100$) making it an ideal comparison sample to our KMOS stars.
We convolved the high-resolution SN sample with a Gaussian kernel to the resolving power of KMOS (R=4000), assuming a constant R over the full wavelength range. We have  run STARKIT  over these degraded spectra using the same setup as mentioned above (see Sect.~\ref{Analysis})

\begin{figure}[!htbp]
\centering
\includegraphics[width=0.49\textwidth]{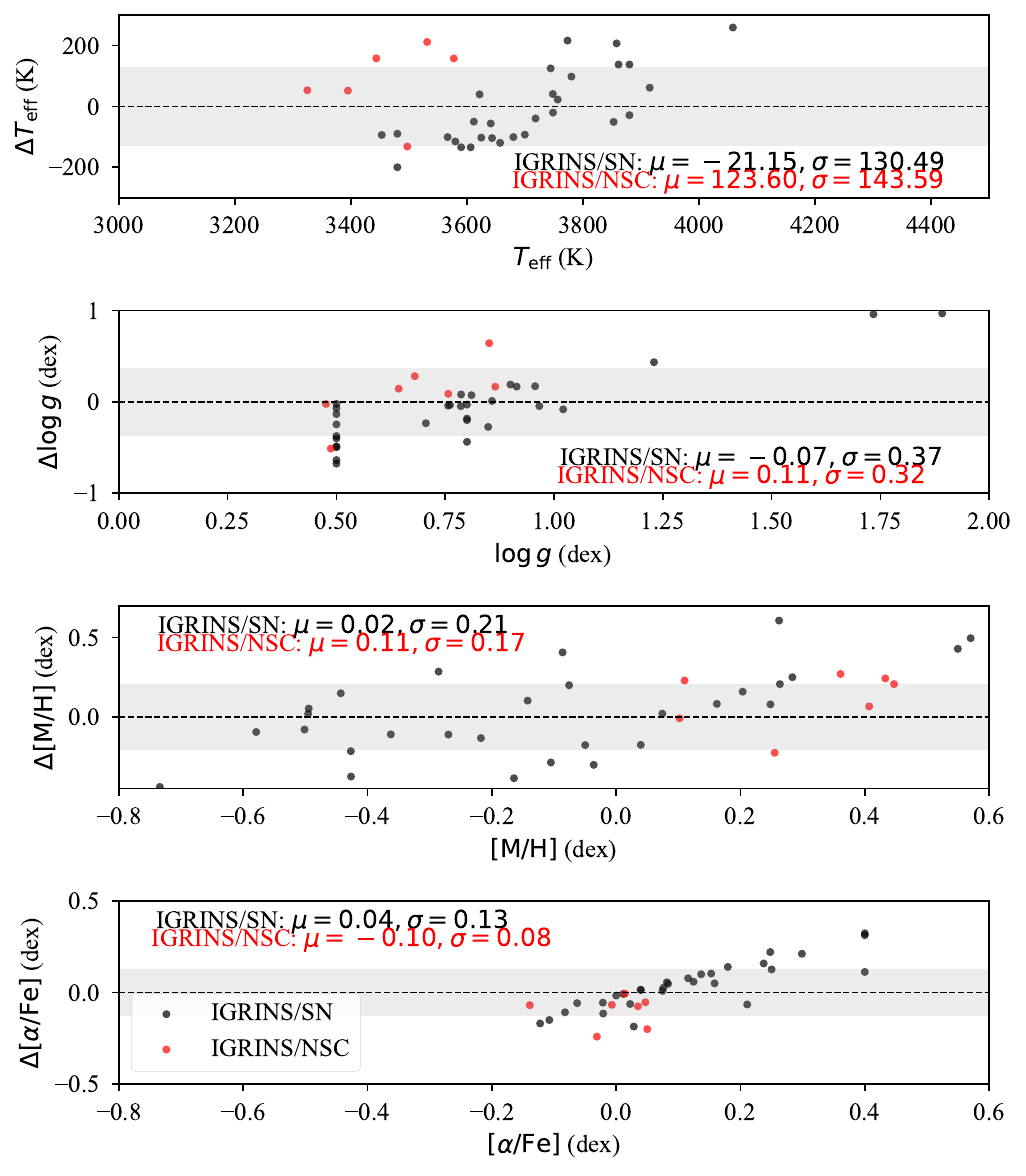}
\caption{Comparison between stellar parameters of high-resolution spectra to degraded spectra at the resolution of KMOS. The y-axis shows the difference between low-resolution and high-resolution spectra, respectively while the x-axis displays the value from the low-resolution work. The shaded gray are shows $\rm \pm 1\,\sigma$ levels for each parameter comparison. Mean difference and standard deviations are indicated on the upper right. Black dots show the SN sample while red dots the NSC stars from \citet{Nandakumar2025}}
\label{stellar_parameters-comparison}
\end{figure}

In addition, seven M giants  in our KMOS sample have been observed with  IGRINS and analyzed  in \citet{Nandakumar2025} and \citet{Ryde2025}. These stars were observed with the same instrumental configuration as the SN sample and can be considered as bench-mark stars in terms of stellar parameters and chemical abundances. We refer to this as IGRINS/MWNSC  sample and refer for a more detailed discussion about the analysis to \citet{Nandakumar2025}.

Figure~\ref{stellar_parameters-comparison} shows the comparison of the stellar parameters ($\rm T_{eff}$, $\rm log\,g$, $\rm [M/H]$, $\rm [\alpha/Fe]$, from upper panel to lower panel). [M/H] is defined as the  global metallicity:  $\rm [M/H] = [Fe/H] + [\alpha/Fe]$.
The black  filled circles show the IGRINS SN sample and the red filled circles the corresponding IGRINS/MWNSC  sample.
In general we see that the stellar parameters can be well recovered from the low-resolution spectra with nearly no systematic offsets. The typical uncertainties are $\rm \sim 150\,K$ in $\rm T_{eff}$, $\rm \sim 0.4 \,dex$ in $\rm log\,g$, $\rm \sim 0.2\, dex$ in $\rm [M/H]$ and $\rm \sim 0.10\,dex$ in $\rm [\alpha/Fe]$. We see some trend in the comparison with the global $\rm \alpha$-values, in the sense that the low-resolution spectra seem to  overestimate the  global $\rm \alpha$ for high $\rm \alpha$-abundances. We have performed a linear fit (last panel of Fig.~\ref{stellar_parameters-comparison}) which gives us  $\rm \Delta [\alpha/Fe] = 0.5814 \times [\alpha/Fe] - 0.0233$ and we applied this correction to our $\alpha$-measurements.    

Nevertheless, Figure~\ref{stellar_parameters-comparison} clearly demonstrates that the stellar parameters can be reasonably well recovered from our KMOS sample.

\begin{figure*}[!htbp]
\centering
\includegraphics[width=0.49\textwidth]{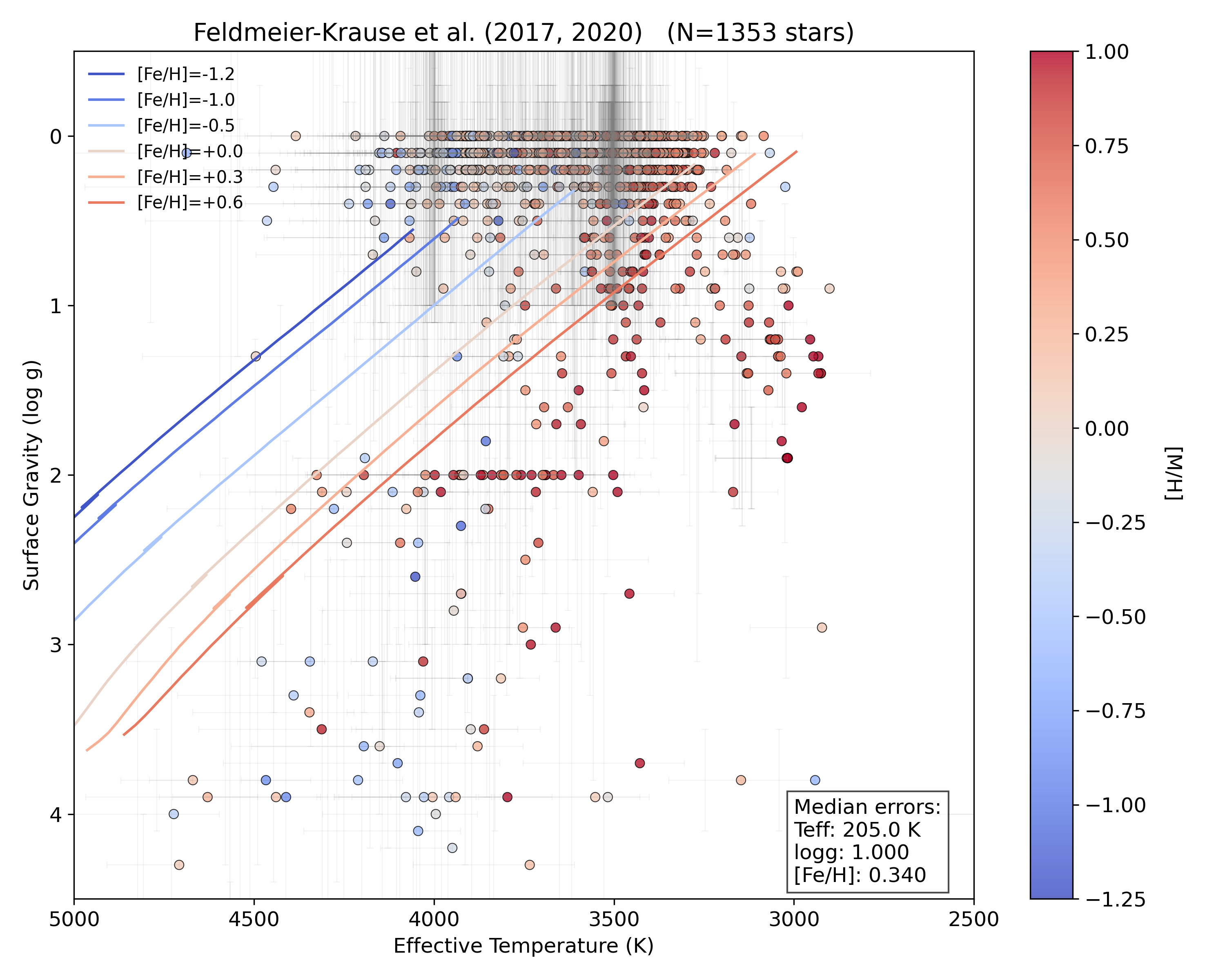} \includegraphics[width=0.49\textwidth]{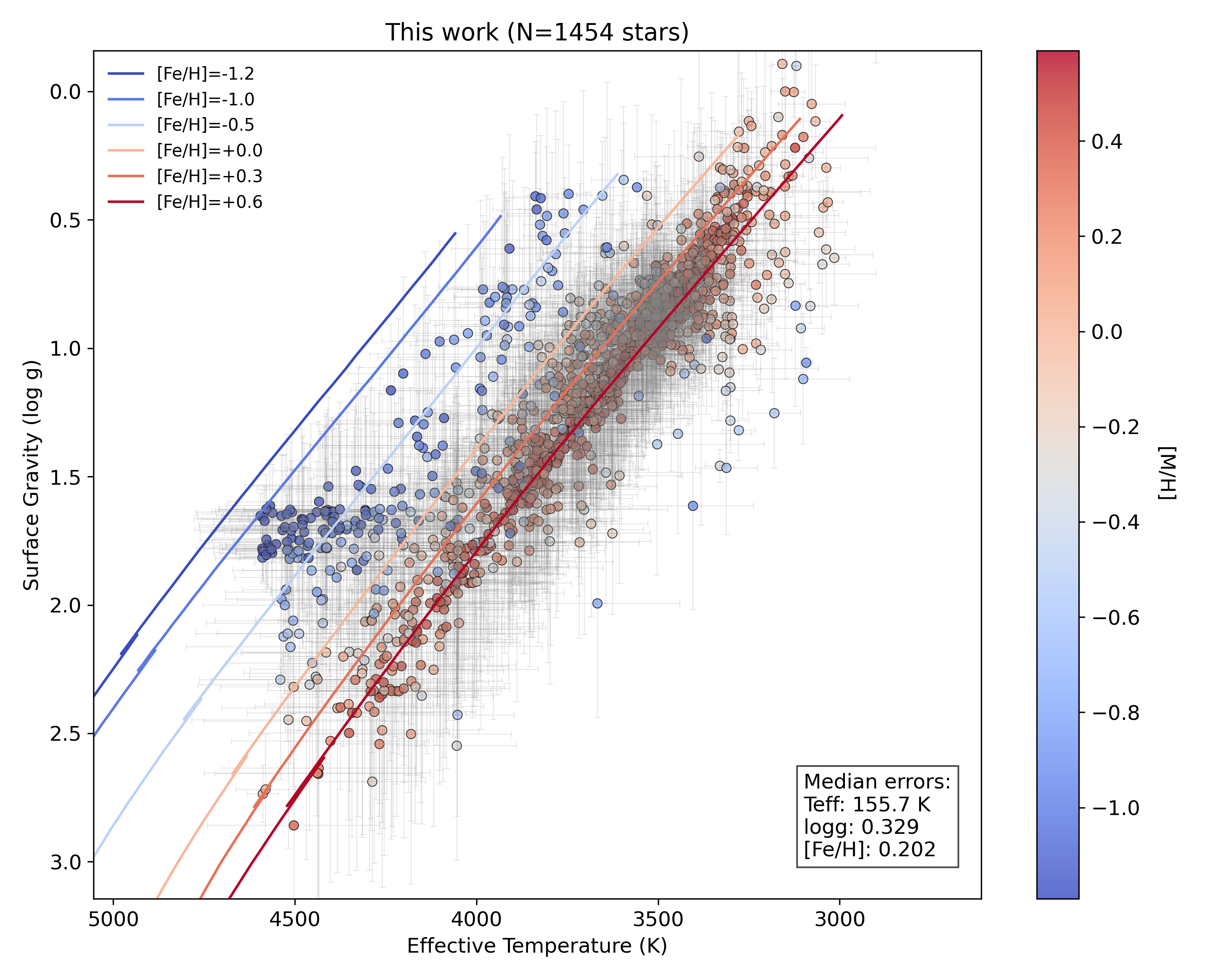}
\caption{Comparison of the Kiel diagram between FK1720 (left panel) and our work (right panel). Superimposed are the PARSEC stellar isochrones with an age of 8\,Gyr and ranging in [M/H]  from -1.2\,dex (dark blue) to  0.6\,dex (dark red). The standard deviations of $\rm T_{eff}$ and $\rm log\,g$ is indicated for each star in grey.}
\label{Teffvslogg}
\end{figure*}

\section{Results}  
Figure~\ref{Teffvslogg} shows the comparison of the Kiel diagram with respect to FK1720 (left panel). We clearly see the improvement in the stellar parameters ($\rm T_{eff}$, $\rm log\,g$, and $\rm [M/H]$) covering the expected parameter space for typical  red giant branch stars which are also indicated by the PARSEC isochrones. Indicated are also the uncertainties for each of our stars. The median uncertainties are in the order of $\rm T_{eff} \simeq 150\,K$, $\rm log\,g \simeq 0.3\,dex$, and $\rm [M/H] \simeq 0.2\,dex$ which are smaller when comparing the uncertainties with FK1720 ($\rm \sigma_{Teff} \simeq 212\,K $, $\rm \sigma_{log\,g} \simeq 1\,dex $, $\rm \sigma_{[M/H]} \simeq 0.26\,dex $). 
Note also that the highest metallicity in our grid is 0.6\,dex and the coolest temperature is restricted to 2800\,K. The grid of FK1720 spans a wider range with temperatures down to 2300\,K and [M/H] up to 1\,dex.  The main reason for this improvement is related to the new grid of synthetic models together with imposing a uniform prior in $\rm T_{eff}$ and $\rm log\,g$.
A slight improvement comes also in the continuum normalization. While FK1720 did the normalization within STARKIT using a $\rm 5^{th}$ order polynomial, we perform the continuum normalization before running STARKIT with a $\rm 6^{th}$ degree polynomial, and applying a sigma clipping (see Sect.~\ref{Analysis}) to remove outliers.

\subsection {$\rm   \alpha$-elements}
The $\alpha$-elements have been obtained by running STARKIT in a second time fixing  the stellar parameters, $\rm T_{eff}$ and $\rm log\,g$, but using [M/H] and $\rm \alpha$ as a free parameter. As in the first step we use all the posterior parameters to trace the uncertainties as well as the border flags.  For our prior we assume that the $\rm \alpha$- elements follow the Galactic prior, meaning that for sub-solar metallicities, the $\rm \alpha$-elements are in the range $\rm -0.2 < [\alpha/Fe] < 0.4$
while for super-solar metallicities, $\rm \alpha-elements$ are solar or sub-solar ($\rm -0.2 < [\alpha/Fe] < 0.2$).

\begin{figure}[!htbp]
\centering
\includegraphics[width=0.49\textwidth]{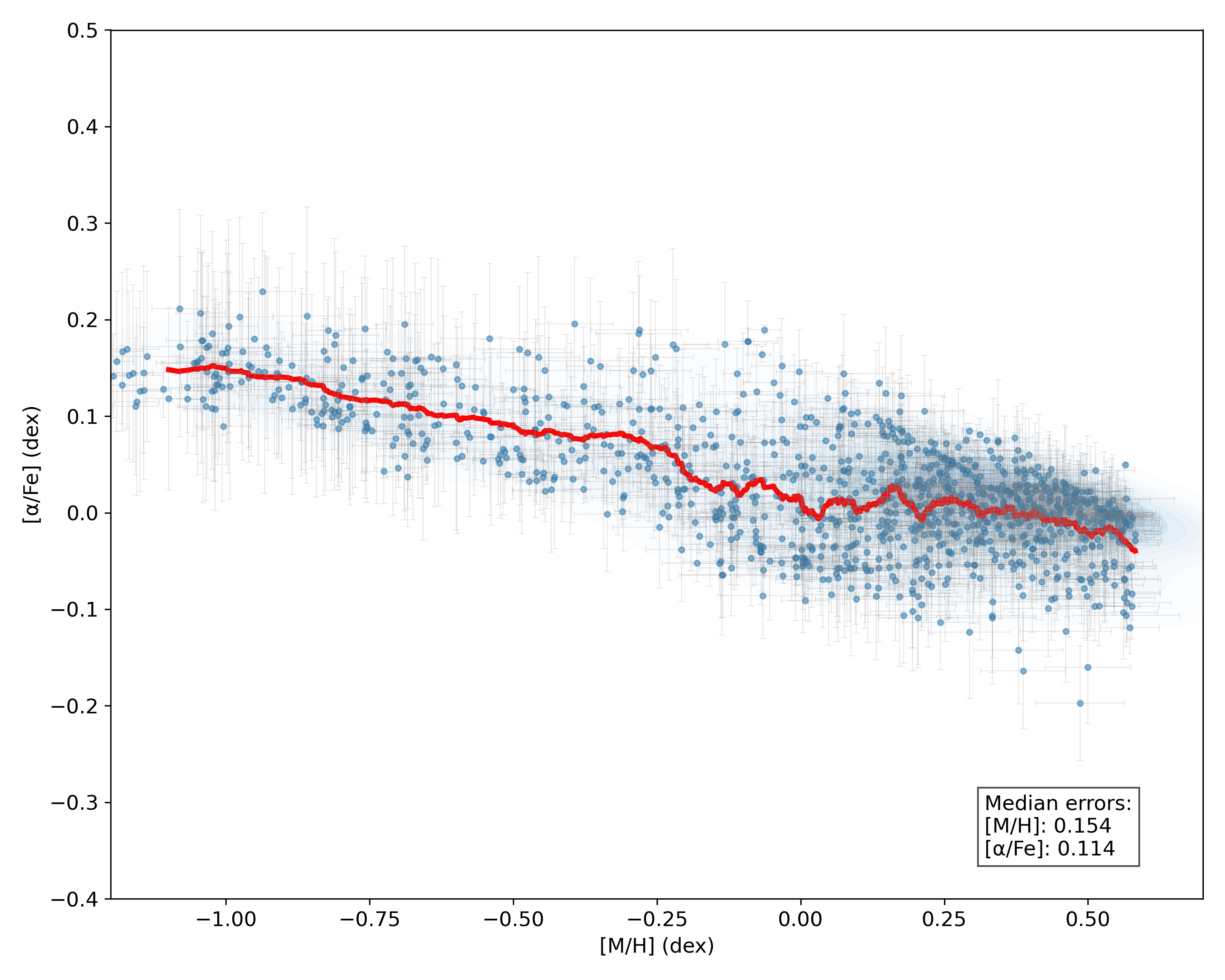}
\caption{$\rm [\alpha/Fe]$ vs.\ $\rm [M/H]$ diagram of 1171 stars. The standard deviations are indicated in grey, while the red line shows the running mean.}
\label{alpha}
\end{figure}

A running mean of $\rm [\alpha/Fe]$, calculated with a dynamic window size (5\% of the dataset or a minimum of 10 stars), was applied to trace the underlying trend across metallicity. Error bars reflect the root-mean-square uncertainties of the averaged $\alpha$-abundances. Figure~\ref{alpha} shows the expected $\alpha$-enhancement at low metallicities, consistent with the enrichment by core-collapse supernovae at early times (\citealt{Matteucci2021}).
We want to stress, that while the global $\alpha$-trend can be seen as representative for the global chemical evolution trend, the individual  measurement uncertainties can be very large and should be treated with caution. This can be seen also in Fig.~\ref{stellar_parameters-comparison} where for high $\alpha$-enhanced stars, the low-resolution spectra overestimate the $\alpha$-abundances with respect to the high-resolution spectra. 

\subsection{Metallicity distribution function}
We analysed the metallicity distribution function (MDF) of our sample by adapting a Gaussian Mixture modelling (GMM) using the Bayesian Information Criterion (BIC) to determine the optimal number of components. The GMM parameters are refined using a MCMC sampler to account for metallicity measurement errors. Posterior distributions from the MCMC chains are used to compute median parameter values and their uncertainties. We generate 1000 random samples from the posterior to construct confidence intervals (68\%, 95\%, and 99\%) for the total metallicity distribution, accounting for both model and measurement uncertainties. The GMM models favors a two component model with a mean metallicity of $\rm -0.77\,dex$ and a sigma of  0.24\,dex for the metal-poor population and +0.26\,dex with a sigma of 0.10\, dex for the metal-rich population.

\begin{figure}[!htbp]
\centering
\includegraphics[width=0.49\textwidth]{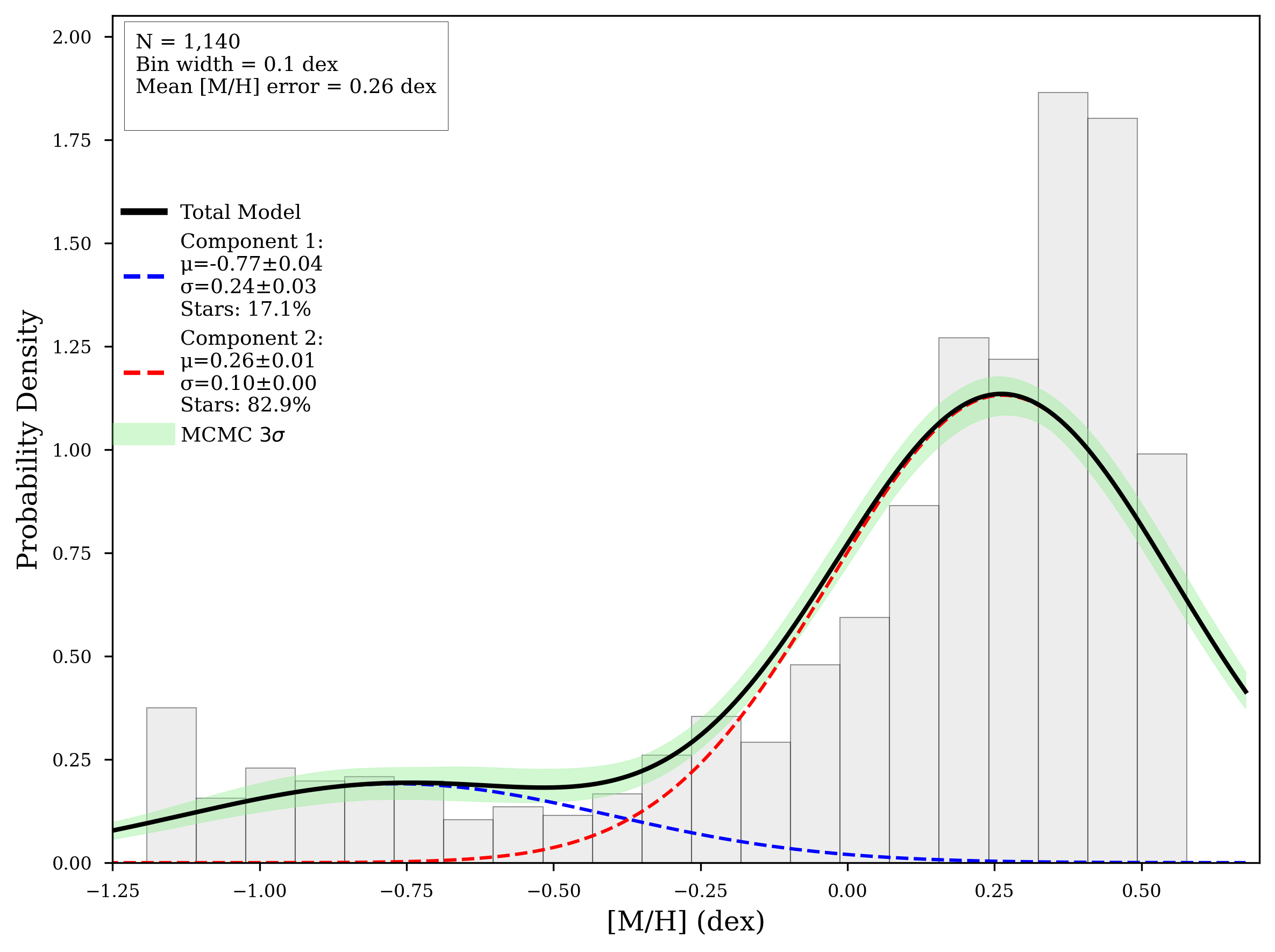}
\caption{ Metallicity distribution function of our sample. A two-component GMM model is superposed. The first component is centered at $\rm -0.76\,dex$ with a std of 0.24\,dex, while the second component is centered at +0.26\,dex with a std of 0.10\,dex. The blue dashed lines represents the Gaussian fit for the metal-poor population while the red dashed lines for the metal-rich population. The total model is displayed with a solid black line together with the $\rm 3\,\sigma$ uncertainties.}
\label{GMM}
\end{figure}

Our findings of a dominant metal-rich population is in agreement with FK1720. However, 
we find  a more significant fraction of metal-poor stars ($\rm \sim 17\%$ compared to 10\% in FK1720) in the MWNSC. \citet{Nogueras-Lara2022} showed that by using extinction values one can separate stars from the NSC and the NSD in the sense that stars in NSC show systematically higher interstellar reddening  values (see Sect. ~\ref{membership}). They concluded that the original FK1720 sample of stars show a significant contribution of stars situated in the MWNSD. They obtain a larger fraction of metal-poor stars compared to FK1720, but their metallicity peaks at around -0.2\,dex which is much more metal-rich compared to our derived values (-0.76\,dex). This difference is certainly due to our reanalysis of the FK1720 sample as we use their same selection criteria to remove stars from the MWNSD.

\begin{figure*}[!htbp]
\centering
\includegraphics[width=\textwidth]{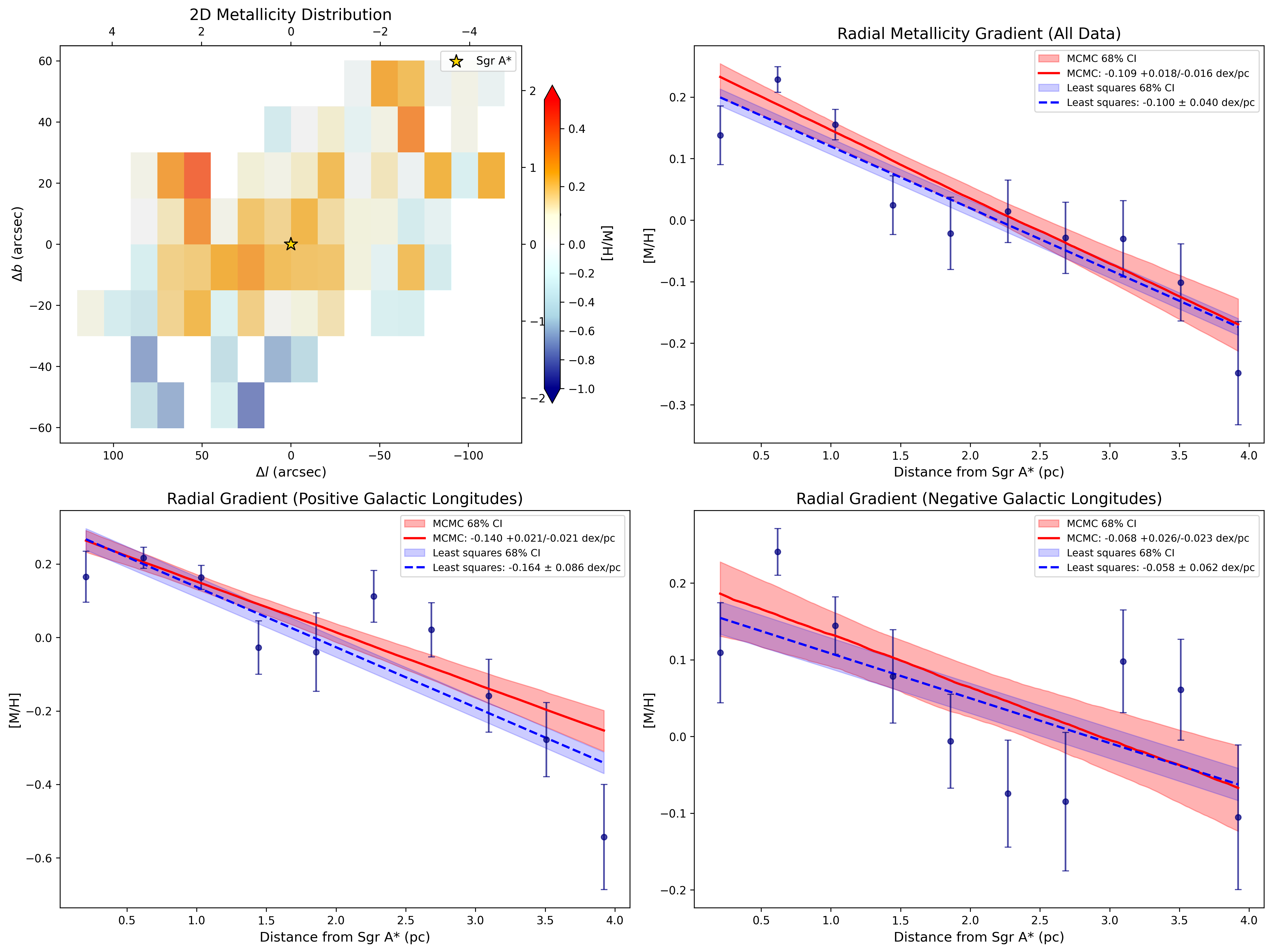}
\caption{(Upper left panel) Mean metallicities in offset coordinates from Sgr~A$^*$. Galactic North is up. Each of the bin has at least 5 stars. (Upper Right panel): Mean metallicity as a function of the distance from Sgr~A$^*$ in pc. Error bars indicate the standard error of the mean metallicity within each bin ( each the bin has at least 5 stars). The blue dashed line shows the a linear least square fit with the 95\% confidence interval shown as the blue area. The red lines show the Bayesian MCMC analysis with the median prediction from the MCMC samples and the red area indicates   the $\rm 16^{th}$-$\rm 84^{th}$ percentile uncertainty range from the MCMC posterior predictions. (Lower  left panel): Similar as upper right panel but only for positive galactic longitudes. (Lower right panel): Similar as upper right panel but only for negative galactic longitudes.}
\label{metallicity_map}
\end{figure*}

\subsection{Metallicity maps and gradients}

As in FK1720, we trace the spatial distribution of the stars in our data set as a function of metallicity to see any possible asymmetries. Figure \ref{metallicity_map} shows on the upper left  panel a metallicity map centered on Sgr~A$^*$. The  bins are chosen to be  15\arcsec\,  wide allowing to have enough stars ($\rm > 5$ stars) in each metallicity bin.

 The resulting metallicity map reveals a dominant  very metal-rich population but we also notice also a  significant high fraction of sub-solar metallicity towards the South. The upper right panel of Fig.\ref{metallicity_map} displays the mean metallicity as a function of the  distance to Sgr~A$^*$, where we see a clear indication of  a negative radial metallicity gradient with $\rm 0.1\,\pm 0.040\,$ dex/pc. We have tested the statistical significance of the metallicity gradient by applying a MCMC analysis using 32 walkers for 5000 steps following a 1,000-step burn-in period. Our MCMC results are highly consistent with the linear regression results with a slope of $\rm -0.109^{+0.018}_{-0.016}\, dex/pc $ with a highly significant p-value of 0.0002 ($\rm R^{2} \sim 0.84$) (see Fig.\ref{metallicity_map}, right panel).  This statistical analysis make us confident of the reliability of our results showing a strong metallicity gradient within the MWNSC. The lower left and lower right panel shows the radial gradient for positive and negative galactic longitudes, respectively where we see some slight impact on the asymmetry at negative galactic longitudes.

\section{Discussions}

\citet{Feldmeier-Krause2022} studied the metallicity distribution of two fields along the Galactic plane, one situated $\rm \sim 20\,pc$ East and West from Sgr~A$^*$ and found a continuous decrease in metallicity with increasing distance with respect to Sgr~A$^*$.
\citet{Nogueras-Lara2023} analyzed the transition region between the MWNSC and the MWNSD to better characterize both structures. They use interstellar reddening as a  proxy of the galactocentric distance  and confirmed the metallicity gradient from \citet{Feldmeier-Krause2022}.
Our  derived metallicity maps and metallicity gradient of the NSC (see Fig.~\ref{metallicity_map})  is consistent with an inside-out formation scenario of the MWNSC.

The formation of the NSD in barred galaxies is mainly due to the  bar-driven gas inflow to the centre of the galaxies (\citealt{review_paper}). The gas inflow rates generate disc or ring like structures, such as the central molecular zone in the Milky Way (\citealt{Morris1996}). The current idea is that star formation occurs in these gaseous rings which build up nuclear stellar discs over time (\citealt{review_paper}). While most of the gas is stalled at the nuclear ring, a small fraction can be transferred further to the  NSC via nuclear inflows to the innermost few pc (see e.g. \citealt{Moon2023}, \citealt{Tress2024}), creating small circum-nuclear rings. These circum-nuclear rings can then grow  with time leading to a metallicity gradient, typical for an inside-out formation scenario.  The Milky Way hosts a small circum-nuclear ring with a radius of 4\,pc (\citealt{Gallego-Cano2020}), which is the typical size of the MWNSC.

Another way to connect the MWNSC and the MWNSD is if the MWNSD starts forming at the centre under the condition that the initial central mass concentration   when the bar has formed is small (\citealt{review_paper}). In that case the bar-driven inflow will drive gas directly to the centre forming the MWNSC and MWNSD at the same time  where the MWNSC is the continuous inner part of the MWNSD. This scenario has been suggested by \citet{Nogueras-Lara2023} predicting the observed metallicity gradient and suggesting that the MWNSC and MWNSD are essentially the same structure, where the MWNSD is the growing edge of the MWNSC.

Concerning NSCs, several possible formation scenarios for in-situ formation have been proposed such as bar-driven gas infall, dissipative nucleation, tidal compression or magneto-rotational instability (see \citealt{Neumayer2020} for more details). The presence of a young stellar population concentrated at small radii (e.g $\rm < 0.5$ pc in the Milky Way, \citealt{Feldmeier-Krause2015}) favours the scenario of an in-situ formation of the NSC which has been also detected in other NSCs such as M31 (\citealt{Carson2015}).

In our work, we establish within the  MWNSC  a negative metallicity gradient, a clear signature of an inside-out formation, in a similar way as the MWNSD. The metal-poor  stars are mainly located in the southern  outer parts of the MWNSC, contrary to \citet{Feldmeier-Krause2020} where they found an anisotropy of low-metallicity stars in the Galactic North. While our fraction of metal-rich stars ($\rm [M/H] > 0.3\,dex$) is 0.64 and similar to \citet{Feldmeier-Krause2020}, we get a much higher fraction of metal-poor stars (17\% compared to 6\% in FK20). We believe that our improved synthetic grid is responsible for this large difference. We also note, as shown in Figure~\ref{alpha}, that we see an increase in $\rm \alpha$-abundances for the high metallicity stars, replicating the results from \citealt{thorsbro2020}.

Unfortunately the datasets of the MWNSC (FK1720, \citealt{Feldmeier-Krause2022}, the MWNSD (\citealt{Schultheis2021}, \citealt{Fritz2021}) have been so far not analysed in the same way. \citet{Fritz2021} uses spectral indices to estimate metallicities based on the equivalent widths of the NaI and the CaI lines while in FK1720 and in this work the full stellar parameter fitting has been used. A full consistent analysis of the full MWNSC and MWNSD covered by the KMOS spectra is essential to draw any further conclusions on the possible interplay between MWNSC and MWNSD.

\section{Conclusions}

We have reanalysed the M giant sample situated in the MWNSC of FK1720 by adopting a new model grid of synthetic spectra using an accurate line list for atoms and molecules and taking into account NLTE effects in our modeling. This line-list has been already successfully tested in high-resolution spectroscopic studies (see e.g. \citealt{Nandakumar2024}, \citealt{Thorsbro2023}, \citealt{Ryde2025}). In addition, we added some  improvements, such as performing a continuum normalization before feeding the spectra to STARKIT, and removing stars which are too close to the border grid of the synthetic spectra. We also obtain $\alpha$-elements from our sample.
A comparison with high-resolution infrared spectra (IGRINS/GEMINI) shows  typical uncertainties of  $\rm \sim 150\,K$ in $\rm T_{eff}$, $\rm \sim 0.4 \,dex$ in $\rm log\,g$, $\rm \sim 0.2\, dex$ in $\rm [M/H]$ and $\rm \sim 0.10\,dex$ in $\rm [\alpha/Fe]$.
Our resulting Kiel diagram shows the expected parameter space for RGB stars and matches the expected trends in metallicity by projecting PARSEC isochrones. 

Our final sample consists of 1140 stars from which we derived the metallicity distribution function. We eliminated contaminated stars from the NSD and the foreground by applying a colour cut in H--K as discussed in \citet{Nogueras-Lara2023}
By applying a GMM model, we identify two populations, one metal-rich centered at $\rm [M/H] \simeq 0.26\,dex $, and another metal-poor one centered at $\rm [M/H] \simeq -0.77\,dex $. We find a higher fraction of metal-poor stars ($\sim 17\%$) compared to FK1720 which are mostly located at the Southern outer parts of the MWNSC. We find a negative radial metallicity gradient of $\rm 0.1 \pm 0.02\,dex/pc$ as a function of the the distance to Sgr~A$^*$ indicating a possible inside-out formation scenario for the MWNSC. 

In the upcoming near future, the Multi-Object Optical and Near-IR Spectrograph (MOONS, \citealt{Cirasuolo2020}) is the forthcoming third-generation instrument to be installed  in 2025 at ESO's VLT in Chile. MOONS will have a spectral resolving power R between 4,000 to 20,000, excellent multiplex capabilities (allowing for the spectra of a thousand objects to be registered simultaneously), and near-IR wavelength coverage (H-band).  MOONS at the VLT will be a unique facility for measuring accurate radial velocities, metallicities and chemical abundances for several million stars across the Milky Way at high spectral resolution. This makes MOONS the ideal instrument to observe stars in the highly-obscured regions of the inner Galaxy such as the MWNSC and the MWNSD.

\begin{acknowledgements}
MS wants to thank the Laboratory of Lagrange (UMR 7293) for the financial support via the BQR. LS thanks the Observatoire de la Côte d'Azur and the Université Côte d'Azur for making this project possible. BT acknowledges the financial support from the Wenner-Gren Foundation (WGF2022-0041). FNL acknowledges support from grant PID2024-162148NA-I00, funded by MCIN/AEI/10.13039/501100011033 and the European Regional Development Fund (ERDF) “A way of making Europe”, as well as from the Severo Ochoa grant CEX2021-001131-S, funded by MCIN/AEI/10.13039/501100011033. A.F.K. acknowledges funding from the Austrian Science Fund (FWF) [grant DOI 10.55776/ESP542]. KF and MCS acknowledge financial support from the European Research Council under the ERC Starting Grant “GalFlow” (grant 101116226)

\end{acknowledgements}

\bibliography{nsc.bib}

\end{document}